\documentclass[prl,aps,twocolumn,showpacs,preprintnumbers,amsmath,amssymb]{revtex4}

\usepackage{graphicx}
\usepackage{dcolumn}
\usepackage{bm}

\begin{document}


\title{Electromagnetically Induced Transparency\\in optically trapped rubidium atoms}

\author{Bernd Kaltenh\"{a}user}
 \email{b.kaltenhaeuser@physik.uni-stuttgart.de}
\author{Harald K\"{u}bler}
\author{Andreas Chromik}
\author{J\"{u}rgen Stuhler}
\author{Tilman Pfau}

\affiliation{5. Physikalisches Institut, Universit\"{a}t Stuttgart, 70550 Stuttgart, Germany.}
\homepage{http://www.pi5.uni-stuttgart.de}

\author{Atac Imamoglu}
\affiliation{Institute of Quantum Electronics, ETH-Z\"{u}rich,
8093 Z\"{u}rich, Switzerland}


\begin{abstract}
We demonstrate electromagnetically induced transparency (EIT) in a
sample of rubidium atoms, trapped in an optical dipole trap.
Mixing a small amount of $\sigma^-$-polarized light to the weak
$\sigma^+$-polarized probe pulses, we are able to measure the
absorptive and dispersive properties of the atomic medium at the
same time. Features as small as 4\,kHz have been detected on an
absorption line with 20\,MHz line width.
\end{abstract}

\maketitle

\section{Introduction}
Electromagnetically induced transparency allows for the
elimination of absorption in an otherwise opaque medium
\cite{Harris1990}. The effect is based on a third state, which is
coupled to the excited state by an additional laser, such that all
possible absorption paths destructively interfere. On the level of
single excitations, the corresponding collective excitations can
be described as a quasi-particle, the so-called dark state
polariton \cite{Fleischhauer2000}. Recently, the particle nature
of dark state polaritons has been experimentally
demonstrated \cite{Karpa2006}.\\
In quantum information processing, photons can be used as robust
information carriers \cite{Knill2001}, but they lack the
possibility of storage. To overcome this shortcoming, several
experiments have used the concept of dark state polaritons to
store photonic information in cold atoms \cite{Liu2001} and vapor
cells \cite{Phillips2001}. Furthermore, as a step towards storage
in a solid, EIT has been demonstrated in a room-temperature solid
\cite{Bigelow2003}. It has also been shown theoretically that EIT
in atomic ensembles can be used to enhance the possibilities of
long-distance quantum communication \cite{Duan2001}.\\
Spin squeezing is often regarded as a benchmark for the control of
atom-light-states. This effect has already been demonstrated in a
vapor cell via a quantum nondemolition measurement
\cite{Kuzmich2000}. In magneto-optically trapped cold atoms it has
been demonstrated in a similar way \cite{Geremia2004}, as well as
by mapping the squeezed state of light onto the atomic
ensemble\cite{Hald1999}. Recently, de Echaniz and co-workers have
shown that this effect can be significantly increased in an optical
dipole trap \cite{deEchaniz2005}.\\\\
Here, we report on the first experimental demonstration of EIT in
an optical dipole trap. Contrary to magneto-optical and magnetic
traps, our setup allows for arbitrary magnetic fields. A
homogeneous magnetic field can be used to
address different magnetic substates of the medium.\\
We have measured 4\,kHz features in the EIT response. This is an
important step towards long storage times of quantum information
in an atomic ensemble and the investigation of trapped darkstate
polaritons.
\section{Experimental Setup}
To prepare an absorbing medium of trapped atoms, we first capture
$4\cdot 10^9$ Rb-87 atoms in a magneto-optical trap. Afterwards,
we apply a Dark-MOT phase (DM) for 25 ms to ensure an efficient
transfer of the atoms to the dipole trap. For the DM, we ramp up
the magnetic gradient field from 13.7 to 18\,G/cm, detune the
MOT-lasers to $-100$\,MHz from resonance and lower the repump
laser power to 1\,\%. After the DM we have $7\cdot 10^8$ atoms at
a temperature of 38\,$\mu$K and a density of
$10^{12}$\,atoms$\cdot\mathrm{cm^{-3}}$
left.\\
The crossed CO$\mathrm{_2}$-laser dipole trap (DT) is turned on
during the loading steps described above. After switching off the
DM and additional 80\,ms thermalization time we capture $2\cdot
10^7$ atoms at a density of $2\cdot
10^{11}$\,atoms$\cdot\mathrm{cm^{-3}}$ in the DT. Due to optical
pumping, the atoms are distributed over the 5 magnetic substates
of the $5S_{1/2},\,F=2$ ground state. Because the potential of the
DT is much steeper than the one of the DM, the cloud heats up to a
temperature of 110\,$\mathrm{\mu K}$. The cloud provides a medium
with an optical density up to $0.76$ for a single Zeeman component
on resonance.\\
\begin{figure}[t!h!c!p]
\begin{centering}
\includegraphics[width=0.5\columnwidth]{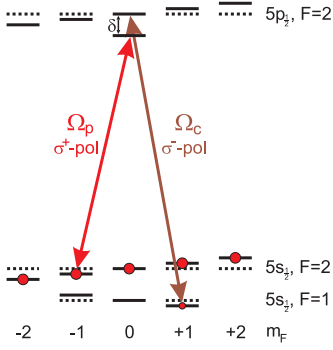}\caption{Level
scheme of the EIT transition. The probe laser couples to the
$5S_{1/2},\,F=2,\,m_F=-1\longleftrightarrow
5P_{1/2},\,F=2,\,m_F=0$ transition, the coupling laser to the
$5S_{1/2},\,F=1,\,m_F=+1\longleftrightarrow
5P_{1/2},\,F=2,\,m_F=0$ transition.}\label{figEIT}\end{centering}
\end{figure}\\
For the EIT-measurements, a magnetic offset field of 129\,G is
applied parallel to the laser beam propagation. At this field
strength, the magnetic substates of the $5S_{1/2}$ ground state
can be addressed individually. This allows to perform the
EIT-measurement only between the ($5S_{1/2},\,F=2,\,m_F=-1$) and
($5S_{1/2},\,F=1,\,m_F=+1$) substates. We use a Raman laser system
to address these transitions, which are shown in figure
\ref{figEIT}.\\
\begin{figure}[h!t!c!p]
\begin{centering}
\includegraphics[width=0.8\columnwidth]{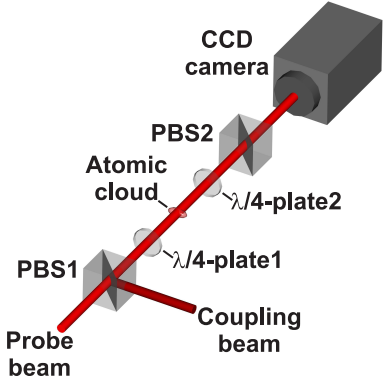}
\caption{Setup for the experiment: the probe and the coupling
laser are overlapped in a polarizing beamsplitter (PBS). With the
following $\lambda/4$-plate the polarization of the pulses are
adjusted before they enter the cloud. With the second
$\lambda/4$-plate the polarizations are turned again to separate
the probe from the coupling beam in the following polarizing
beamsplitter. Due to lenses (not shown in the picture), the cloud
is imaged onto a high efficiency CCD camera.
}\label{figsetup01}\end{centering}
\end{figure}\\
The setup for the EIT-measurements is shown in figure
\ref{figsetup01}. For revealing the dispersive properties of the
medium, the first $\lambda/4$-plate is turned, until a
$\sigma^-$-polarized intensity admixture of $a^2=8.7\,\%$ to the
probe beam is obtained. The second $\lambda/4$-plate compensates
this effect and mixes both polarizations back to linearly
polarized light, which is then measured beyond the second
polarizing beamsplitter. This causes the two polarizations to
interfere. A similar method has recently been demonstrated with a
vapor cell in a Sagnac interferometer
\cite{Purves2006}.\\
Due to the large Zeeman-shift of the magnetic substates, the
Raman-condition is not fulfilled for the wrong polarizations and
thus the admixture in the coupling beam can be neglected.
\section{Theory}\label{theory}
\begin{figure}[h!t!c!p]\begin{centering}
\includegraphics[width=0.5\columnwidth]{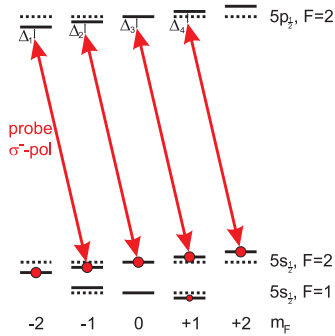}\caption{Level
scheme for the $\sigma^-$-polarized component of the probe light:
the detunings of the respective transitions $j$ are marked as
$\Delta_j$.}\label{figSigmaminus}\end{centering}
\end{figure}

As described above, the first $\lambda/4$-plate mixes a relative
intensity $a^2$ of $\sigma^-$-polarization into the otherwise
$\sigma^+$-polarized probe beam. Due to birefringence in the
optical viewports of the vacuum chamber, the $\sigma^-$-polarized
beam collects an additional phase $\phi$ relative to the
$\sigma^+$-polarized beam. The total electric field acting on the
atoms can then described via
\begin{eqnarray}
\left|E_{in}\right|^2&=&\left|E_{in,\sigma^+}+E_{in,\sigma^-}\right|^2\nonumber\\
&=&\left|\sqrt{1-a^2}E_0+aE_0\exp\{i\phi\}\right|^2\nonumber\\
&=&E_0^2\left(1+2a\sqrt{1-a^2}\cos\phi\right)\;.\label{eq01}
\end{eqnarray}
When we tune the coupling laser to resonance, the single-photon
and two-photon detuning of the probe laser become identical and
the susceptibility for the $\sigma^+$-polarized probe laser is
given by \cite{Fleischhauer2005}
\begin{eqnarray}
\chi^{(+)}=&&\frac{|\mu|^2\varrho}{\epsilon_0\hbar}\nonumber\\
&&\times\left[\frac{4\delta(\Omega_c^2-4\delta^2-\gamma^2)}{|\Omega_c^2+(\Gamma+i2\delta)(\gamma+i2\delta)|^2}\right.\nonumber\\
&&+i\left.\frac{8\delta^2\Gamma+2\gamma(\Omega_c^2+\gamma\Gamma)}{|\Omega_c^2+(\Gamma+i2\delta)(\gamma+i2\delta)|^2}\right]\;.\label{eq02}
\end{eqnarray}
To derive this equation, we have also assumed that the relevant
atomic population stays mainly in the initial
$5S_{1/2},\,F=2,\,m_F=-1$ state. This is fulfilled, if a strong
coupling laser or weak probe pulses ($N_{photons}\ll N_{atoms}$)
are used. Here, $\delta$ is the probe laser detuning, $\Omega_c$
the Rabi-frequency of the coupling laser, $\Gamma$ the spontaneous
emission rate between the excited state and the respective ground
state, $\gamma$ the collisional decay rate between the two ground
states and $|\mu|$ the dipole matrix element between the ground
and the excited state.\\
Due to the large Zeeman-shift, the $\sigma^-$-polarized beam does
not fulfill the Raman-condition and thus its susceptibility can be
described by the two-level atom. As can be seen in figure
\ref{figSigmaminus}, one has to sum over the susceptibilities of
all four independent two-level systems, that can interact with the
beam. Due to the large detuning from resonance, absorption can be
neglected ($<0.04\,\%$ in our system), but the phase shift can
become considerable.
\\
The susceptibility is then described by \cite{Suter1997}
\begin{equation}
\chi^{(-)}=\sum\limits_{j=1}^4\frac{\left|\mu_j\right|^2\varrho_j}{\hbar\epsilon_0}\,\frac{\Delta_j+i\frac{\Gamma}{2}}{(\frac{\Gamma}{2})^2+\Delta_j^2}\label{eq03}\;.
\end{equation}
Here, $\varrho_j$ are the populations in the respective ground
states, $\mu_j$ the dipole matrix elements and $\Delta_j$ the
detunings relative to the respective transition, while the decay
rate $\Gamma$ is the same for all of them. The detunings $\Delta_j$
also depend on the probe detuning $\delta$.\\
The electric output field is then given by
\begin{eqnarray}
\left|E_{out}\right|^2&=&\left|E_{out,\sigma^+}+E_{out,\sigma^-}\right|^2\nonumber\\\nonumber\\
&=&\Big|\sqrt{1-a^2}E_0\exp\{i\chi^{(+)}kz/2\}\nonumber\\
&&+aE_0\exp\{i\phi\}\exp\{i\chi^{(-)}kz/2\}\Big|^2\nonumber\\\nonumber\\
&=&a^2\exp\{-\mathrm{Im}\chi^{(-)}kz\}\nonumber\\
&+&(1-a^2)\exp\big(-\mathrm{Im}\chi^{(+)}kz\big)\nonumber\\
&+&2a\sqrt{1-a^2}\exp\left\{-\big(\mathrm{Im}\chi^{(-)}+\mathrm{Im}\chi^{(+)}\big)kz/2\right\}\nonumber\\
&&\times\cos\left\{\phi+\big(\mathrm{Re}\chi^{(-)}-\mathrm{Re}\chi^{(+)}\big)kz/2\right\}\;.\label{eq04}
\end{eqnarray}
It can be seen that the first two terms of the equation describe
the usual behavior, described the respective susceptibility, while
the last term is responsible for the interference and results in
the appearance of the dispersive properties of the medium.\\
Together with equations \ref{eq01} and \ref{eq02}, this yields the
total transmission through the medium via
\begin{equation}
T(\delta)=\frac{\left|E_{out}\right|^2}{\left|E_{in}\right|^2}\;.\label{eq05}
\end{equation}
Because we are probing the sample with relatively short pulses,
the pulse length limits the minimal EIT bandwidth. The Gaussian
pulses are defined as
\begin{equation}
I(t)=I_0\exp\left\{-\frac{t^2}{\tau^2}\right\}\;.
\end{equation}
To include this limitation, one has to evaluate the convolution
integral over the Fourier transformed of the Gaussian pulse
\begin{equation}
F(\delta)=\int\limits_{-\infty}^{+\infty}I(t)\exp\left\{I2\pi\delta
t\right\}dt=\sqrt\pi\tau\exp\left\{-\pi^2\tau^2\delta^2\right\}\;,
\end{equation}
which finally yields the transmission through the cloud:
\begin{equation}
T_P(\delta)=\int\limits_{-\infty}^{+\infty}T(\delta^\prime)F(\delta-\delta^\prime)d\delta^\prime
\end{equation}
Unfortunately, there is no analytic solution to this integral.
\section{Experimental Results}
\begin{figure}[h!t!c!p]
\begin{centering}
\includegraphics[width=0.8\columnwidth]{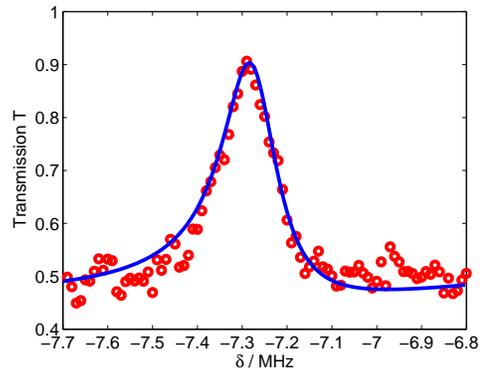}
\caption{Transmission spectrum of a $20\,\mu$s pulse at a coupling
laser Rabi-frequency of 1200\,kHz. The absorptive and dispersive
parts in the signal can be recognized. The frequency offset of
$\delta_0=-7.27\,$MHz corresponds to the differential quadratic
Zeeman shift between the two ground state levels. This offset does
not depend on the lasers and can thus be used to calibrate the
magnetic offset field. For the fit we used equation \ref{eq05} as
an approximation.}\label{figdemo01}\end{centering}
\end{figure}

\begin{figure}[h!t!c!p!b]
\begin{centering}
\includegraphics[width=0.8\columnwidth]{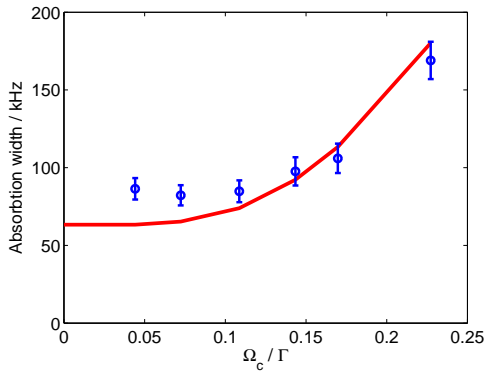}
\caption{Theory curve and EIT measurement with $5\,\mu$s pulses.
The figure shows the transparency width (Gaussian $1/e$-radius)
depending on the Rabi-frequency of the coupling laser. The probe
pulses contain $3\cdot 10^5$ photons within the size of the cloud,
which correspond to a maximum Rabi-frequency of 190\,kHz. The
errorbars reflect the uncertainty in the phase
$\phi$.}\label{figmess01}\end{centering}
\end{figure}

\begin{figure}[h!t!c!p!b]
\begin{centering}
\includegraphics[width=0.8\columnwidth]{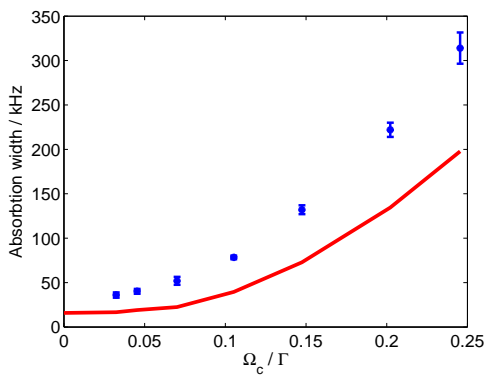}
\caption{Theory curve and EIT measurement with $20\,\mu$s pulses.
The figure shows the transparency width (Gaussian $1/e$-radius)
depending on the Rabi-frequency of the coupling laser. The pulses
contain $2\cdot 10^6$ photons within the size of the cloud, which
correspond to a maximum Rabi-frequency of 220\,kHz. The width is
much narrower than the one of the $5\,\mu$s
pulses.}\label{figmess02}\end{centering}
\end{figure}

\begin{figure}[h!t!c!p!b]
\begin{centering}
\includegraphics[width=0.8\columnwidth]{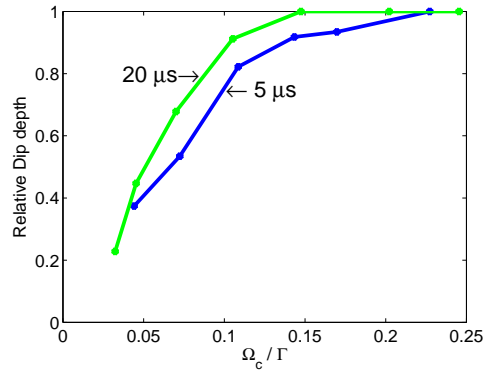}
\caption{EIT measurement with $5\,\mu$s and the $20\,\mu$s pulses.
The figure shows the relative depth of the EIT dip depending on
the Rabi-frequency of the coupling laser. The decrease for small
Rabi-frequencies corresponds to the transient effect that a steady
state cannot be reached here within the time of a short probe
pulse.}\label{figmess03}\end{centering}
\end{figure}

\begin{figure}[t!h!c!p]
\begin{centering}
\includegraphics[width=0.8\columnwidth]{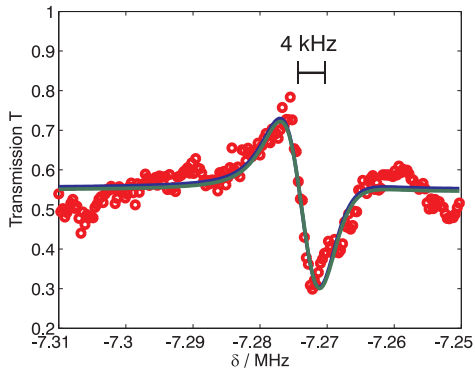}
\caption{EIT measurement with a $100\,\mu$s pulse: the line width
was reduced to 4\,kHz. The transparency is so low that only the
dispersive part of the signal can be recognized. }
\label{figdemo02}\end{centering}
\end{figure}

We have measured the EIT-resonance spectrum for three different
lengths of the probe pulse: $\tau=5\mu$s, $\tau=20\mu$s and
$\tau=100\mu$s. Figure \ref{figdemo01} shows the data of one
measurement with a pulse length of $20\,\mu$s and a coupling laser
Rabi-frequency of 1200\,kHz. In this measurement, it can be seen,
that the signal contains an absorptive (the peak itself) as well
as a dispersive (the asymmetry) part.
\\
The value for the phase $\phi=4.95$ was obtained from the fits of
all measurements. The curve was fitted with equation \ref{eq05}
and yielded $\sigma=100\,$kHz, $\gamma=8$\,kHz and
$\delta_0=7.27$\,MHz for the frequency offset due to the quadratic
Zeeman shift. The ground state decay rate $\gamma$ usually
corresponds to collisions between the atoms as well as collisions
with the background gas. The collision rate rate can usually be
neglected, especially in case of large coupling laser
Rabi-frequencies. But it can also correspond to a transient
effect: for low coupling laser Rabi-frequencies, a steady state in
the atomic population cannot be reached within the time of a short
probe pulse. This effect shows the same empiric behavior as the
collisional loss of polaritons
and leads to non-negligible values of $\gamma$.\\
The data in figures \ref{figmess01} and \ref{figmess02} show the
results of the measurements with the $5\,\mu$s and the $20\,\mu$s
pulses. For large coupling laser Rabi-frequencies, the coupling
laser broadens the line width,
while for lower Rabi-frequencies, the pulse length is the limiting factor.\\
\\
\\
The solid curves show the line width that should in theory be
obtainable with our setup. For large Rabi-frequencies applied on
$5\,\mu$s probe pulses, the measurements are in good accordance
with the theory. For all others, the measured line widths are
broader than the theory for an optical density of $0.76$ predicts.
We attribute these small discrepancies to a decrease in the
optical density of the trapped cloud during the experimental
measurements. Smaller optical densities can be caused by a reduced
number of optically trapped atoms, which is typically observed in
the course of the day, and lead to broader theoretically expected
line widths. The theory curve is plotted for an optical density of
0.76.\\
The lack of sufficient coupling light results in an uncomplete
transparency and limits the relative depth of the EIT dip in the
signal. This can be seen in figure \ref{figmess03}.
\\
To obtain a very narrow line width, a measurement was made with
$100\,\mu$s long pulses, containing $3.9\cdot 10^6$ photons within
the size of the cloud, which corresponds to a maximum
Rabi-frequency of 360\,kHz. Figure \ref{figdemo02} shows the
result for a coupling laser Rabi-frequency of 590\,kHz. For lower
values, the
induced transparency was too low.\\
Due to inefficient EIT, the absorptive part is so low that it is
not visible anymore. Instead, due to a large phase shift, the
dispersive part of the signal gets enhanced, compared to the
measurements shown before.\\
\\
To enhance the dispersive effect, the $\sigma^-$-intensity
admixture $a^2$ was increased to $25$\,\%, which also resulted in
a different differential phase shift $\phi=4.1$. With a Gaussian
$1/e$-half width of 4\,kHz, this is to our knowledge the narrowest
EIT signal measured in ultracold atoms \cite{Hau1999, Braje2004}.
Narrower signals of $\sim 30$\,Hz have been measured in buffer gas
cells, where one is not limited by pulse lengths \cite{Brandt1997,
Erhard2000}.
\section{Conclusion and outlook}
We have shown results on measuring electromagnetically induced
transparency (EIT) in pure optically trapped rubidium atoms. The
signals yield absorptive and dispersive properties of the atomic
medium at the same time. Furthermore, we have measured the
narrowest EIT line width in ultracold atoms.\\
This experiment is an important step towards
polarization-dependent long time storage of quantum information in
an atomic cloud and the investigation of trapped dark state polaritons.\\
In our measurements we are still limited by the relatively low
optical density of $0.7$. The next step will be to optimize the
cooling schemes and therefore increase the optical density. This
will result in an enhanced atom-light interaction, required for
better quantum information processing experiments.
\section{Acknowledgements}
We gratefully acknowledge the support of the Alexander von
Humboldt foundation and the Landesstiftung Baden-W\"{u}rttemberg.
\bibliography{PaperEITinDF01}
\end{document}